\documentstyle[aps,prl,epsfig,multicol]{revtex}
\makeatletter
\newlength\abovecaptionskip \newlength\belowcaptionskip
\def\@makecaption#1#2{%
 \vskip\abovecaptionskip \sbox\@tempboxa{#1: #2}%
 \ifdim \wd\@tempboxa >\hsize #1: #2\par \else \global \@minipagefalse
 \hb@xt@\hsize{\hfil\box\@tempboxa\hfil}%
 \fi \vskip\belowcaptionskip} \makeatother

\def\eps{\varepsilon}
\def\dd{\text{d}}
\def\e{{\text{e}}}
\newcommand\lf{{\lambda _ {\text{\scriptsize{F}}}}}

\begin{document}
 \title {Anderson Orthogonality Catastrophe in Disordered Systems}
 \author{Yuval Gefen$^1$, Richard Berkovits$^2$, Igor V. Lerner$^3$,
 Boris L. Altshuler$^{4,5}$}
\address{ $^1$ Department of Condensed Matter Physics,
    The Weizmann Institute of Science, Rehovot 76100, Israel\\
    $^2$The Minerva Center, Department of Physics,
    Bar-Ilan University, Ramat-Gan 52900, Israel\\
$^3$School of Physics and Astronomy, University of Birmingham,
Birmingham~B15~2TT, United Kingdom\\
$^4$NEC Research Institute, 4 Independence Way, Princeton, New Jersey 08540\\
$^5$   Department of Physics, Princeton University, Princeton, New Jersey
08545 }
    \date{\today} \maketitle
 \maketitle
 \begin{abstract}
 We study the Anderson orthogonality catastrophe (AOC) in finite
conductors with diffusive disorder. The disorder averaged
logarithm of  $\chi$, the overlap between the ground states before
and after adding
  a  static impurity, is found to depend {\it nonmonotonically} on
the disorder.  In two dimensions   $\langle \ln\chi^{-1}\rangle
\propto \ln^2 N$ in the weak disorder limit, thus showing a
stronger dependence on the number of electrons $N$ than in the
canonical AOC. A very broad tail of
 the distribution of $\chi$, found numerically, is a signature of
 the importance of a few-level statistics at the Fermi energy.
 \end{abstract}

\begin{multicols}{2}

 The  Anderson orthogonality catastrophe is ubiquitous in physics,
underlying such disparate phenomena as the zero-bias anomaly in
disordered systems, the fractional Quantum Hall effect and the
metal-insulator transition \cite{AOC_general}. The seminal
paper of Anderson \cite{Anderson} addressed a system of
non-interacting electrons , confined to a finite volume.
No randomness (or complex geometry) were considered. The
introduction of a static impurity modifies the many-particle
ground state (GS) of the original system, such that the overlap
between the original and the modified GS, $\widetilde
\chi=|\langle \Psi |\Psi' \rangle |^2$, is bounded (from above) by
\begin{equation}
 \label{chi} \chi = \exp\biggl[-\frac12\sum_ {\eps_n\le0 ,\, \eps
_{m'}>0 } |\bigl<n|m'\bigr>|^2\biggr]\equiv\exp[-\cal I\,] \, .
\end{equation}
Here $|n\rangle$  and $|m'\rangle$ are the single-particle
 eigenstates of  the
original and the modified system   respectively,
 {with energies $\eps _{n},
\eps _{m'}$ counted from the Fermi level;
 }
 $\cal I$ is referred to
as the Anderson integral. For a short range impurity whose potential %
{has the strength } %
$\lambda $ it was found \cite{Anderson} that
\begin{equation}
\label{ical}
{\cal I}
= 2\pi^2\lambda^2 \ln N,
\end{equation}
where N is the number of electrons: hence a ''catastrophe'' as $N$
approaches the thermodynamic limit.

Since the  AOC appears to be a rather general phenomenon, it is
only natural to study  it under more general circumstances. The
importance of the  inclusion of disorder is self-evident
\cite{chen92,matveev_aliener,smolyarenko_simons,montambaux}. Not
only may the disorder-averaged value of the overlap be different
from its value in a pure system, but it can also induce mesoscopic
fluctuations of ${\cal I}$ which are not necessarily narrowly
distributed. Furthermore, one may consider longer range impurities
as well: in 2DEG systems impurities are located off the conducting
layer, implying softer impurity potentials.  

In this work we consider electronic
systems in the presence of quenched
impurities that give rise to diffusive motion of the electrons,
with an {\it additional}  impurity further introduced into the system.
We address two main
issues 
concerning the AOC caused by the additional impurity.

(i) We consider the average value of the Anderson integral as a
function of $N$, of disorder, characterised by  
the electron transport mean free path, $\ell$,
and of the {\it range}, $b$, of the added impurity 
 parameterized by
\begin{equation}
 \label{u0}
U_0\equiv \frac{2\pi \lambda}\nu u({\bf r}) =\frac{2\pi \lambda}
\nu (\pi b^2)^{-d/2}\e^{-r^2/b^2} \,, \label{imp}
\end{equation}
 where $\nu$ is the single-particle density of states.  We find
that in the weak disorder regime
and for $p_F^{-1}\alt b\ll L$ in 2d ($L$ is the linear system size)  and
$\ell\alt b \ll L $ in 3d \cite{corrections}
\begin{equation}
\langle {\cal I} \rangle= \left({\pi^2\lambda^2}/{g_0}\right) \cases{ \ln^2
N,
 & $d$=2 \cr  (\ell/b) \ln N, & $d$=3\,.} \label{i2}
\end{equation}
Here 
$g_0=2\pi^2\nu D\equiv 2\pi^2\nu\,\upsilon_F\ell/d$ is
the conductance of a $d$-dimensional cube of
size $\ell ^d$. Thus, in 2d the
decrease  with $N$ of the overlap $\chi$, Eq.~(\ref{chi}), 
is faster than a power law, showing a stronger $N$
dependence than in the original AOC for clean systems.  This is
due to the diffusive character of electron motion which increases
the probability of return to the added impurity and thus the
sensitivity of the modified GS
to it. %this impurity.
The overlap decreases faster when disorder increases, while in the
strong disorder regime (localization) it should practically
disappear -- hence  a non-monotonic dependence of $\chi$ on $g_0$.

There exists also a ballistic contribution to the overlap,
\begin{equation}
\langle {\cal I} \rangle= 
 \left({2\pi^4\lambda^2}/{g_0}\right)
\left({\ell}/{\pi b}\right)^{\!d-1} \ln
N\,.\label{i1}
\end{equation}
 For $b \alt p_F^{-1}$ it goes over to the ``clean'' result (\ref{ical}).
 For $p_F^{-1}<b<\ell$  it dominates in 3d, while in 2d,
for the maximal $N$
possible in the weak disorder regime ($\ln N\alt g_0$%%or $\ln N\alt g_0^2$
%for the GOE and GUE symmetries respectively
, see below), the diffusive contribution (\ref{i2}) prevails.

(ii) We observe the appearance of a wide distribution,
$\rho(\chi)$, of the GS overlaps. Numerically we find it to be
anomalously broad even on a logarithmic scale. We {argue} that the
tails of the distribution are defined by the level statistics of
the last occupied and the first unoccupied levels. By projecting
the problem onto a two-level Hilbert space and using random matrix
theory, we derive this distribution non-perturbatively and show it
to be in  good agreement with the numerical results.  The
underlying physics of this part is closely related to a recent
study of the AOC in the context of parametrically varied random
matrices \cite {Vallejos}.

The logarithm of the overlap integral, $\cal I$, can be exactly expressed in
terms of Green's functions:
\begin{equation}%*******************************************************%
 \label{I}%**********************************************************%
{\cal I}=\frac{\lambda^2}{2\nu^2} \int\!\dd {\bf r}\, \dd {\bf r} '
\biggl\{\!u({\bf r}) u({\bf r}') \!\! \int\limits_{-\infty}^{0^-}\!\!\dd
\eps\int\limits_{0^+}^{\infty}\!\!\dd \eps' \frac{{g^* (\eps) g(\eps') }}
{(\eps\!-\!\eps'\!-\!  i0)^2}\biggr\},
\end{equation}%***********************************************************%
where $g (\eps)\equiv G^R(\eps;{\bf r} , {\bf r} ')- G^A(\eps;{\bf r} ,
{\bf r} ')$. Scattering from all the other impurities is implicitly
included in the formally exact retarded and advanced Green's functions,
$G^R$ and $G^A$. The presence of disorder does not have a significant
impact on the {\it mean} value of the overlap ${\cal I}$ provided that the
additional impurity is point-like i.e.\ $b\alt \lf$ in Eq.~(\ref{u0}).
 Taking the limit $b\to0$, i.e.\ $u(\bf r)
=\delta (\bf r)$, one reduces the overlap logarithm (\ref{I}) to
\begin{equation}%**********************************************************%
\label{Int2}%**************************************************************%
 {\cal I}=\frac{(2\pi\lambda)^2}{2\nu^2} \int\limits_{-\infty}^{0^-}\dd
\eps\int\limits_{0^+}^{\infty}\dd \eps' \frac{\nu(\eps) \nu(\eps') }
{(\eps-\eps'-i0)^2}\ .
\end{equation}%************************************************************%
Then, in the leading order in $1/g_0$, the average $\left<\nu\nu\right>$
is given by the disconnected contribution, $\left<\nu\right>
\left<\nu\right>$.
 Approximating $\left<\nu(\eps)\right>$ by  $\nu$ for
$|\eps|<E/2$ ($E$ is the bandwidth) and vanishing outside the
band, one finds the average of (\ref{Int2}) equal to the Anderson
expression, $\left<{\cal I}\right>=2\pi^2\lambda^2 \ln
(E/\Delta)$, where $\Delta = 1/\nu L^d$ is the mean
single-particle level spacing.

When the range of $u({\bf r})$  is finite,
 the mean value of ${\cal I} $ is more strongly affected by disorder.
 Now one needs to take into account both the disconnected (short-range),
$\left<g^*(\eps)\right> \left<g(\eps')\right>$, and connected
(long-range), $\left<g^*(\eps)g(\eps')\right>_{\text{c}}$, contributions.
The former is given by \cite{mirlin,footnote_e}
\begin{equation}%*************************************************************
\langle g(\varepsilon)\rangle \langle g(\varepsilon')\rangle = (2
\pi \nu)^2 e^{-R/\ell} \cases{ J_0^2(k_F R) ,& $d$=2 \cr\cr
\displaystyle \frac{\sin^2(k_F R)}{(k_F R)^{2}},& $d$=3,}
\label{gg}
\end{equation}%***************************************************************
where $R  = |{\bf R}|\equiv|{\bf r} - {\bf r'}|$. Substituting
Eq.\ (\ref{gg}) into Eq.\ (\ref{I}), and performing the spatial
integration  and subsequently the energy integration yields Eq.\
(\ref{i1}).

The long-range contribution is obtained from Eq.\ (\ref{I}) by considering
the impurity average over the advanced-retarded parts of $g^*g$ yielding
\begin{eqnarray}%***************************************************************
\langle {\cal I}\rangle=2\pi\lambda^2\Re{\rm e}\!\!
  \int _\Delta^{1/\tau}\!\! \frac{\dd \omega}\omega{{\Delta}}
\sum_{\bf q}  \int \dd{\bf r} \dd{\bf r'} {\frac{ u({\bf r}) u({\bf
r'})\,\e^{i {\mathbf q} \cdot {\mathbf R}}}{D q^2 - i \omega}} \ .
\label{long}%***************************************************************
%\nonumber
\end{eqnarray}%***************************************************************
Substituting the expression for $u$, Eq.\ (\ref{imp}), performing the
spatial integration, replacing the summation over ${\bf q}$ by the
integration and finally performing the integral over $\omega$ we obtain
Eq.~(\ref{i2}), where we have used $\ln (L/b)^2=\ln
N-\ln[(k_Fb)^2/4\pi]\approx \ln N$ for $b\ll L$ \cite{long-ranged}. The
incipient AOC becomes negligibly small when the impurity is delocalized
over the entire system.

We have thus found that once the impurity range exceeds $p_F^{-1}$ in
$d=2$ or $\ell\gg p_F^{-1}$ in $d=3$, the main contribution to the overlap
integral is the long-range term (\ref{i2}). In $d=2$, it gives rise to the
suppression of $\chi$ which is {\it stronger} than a power law in $N$.
Furthermore, the
 condition that the magnitude of this effect exceeds   the
original AOC factor (\ref{ical})  is
\begin{equation}
L> \ell \exp{g_0}\,. \label{l1}
\end{equation}
At the same time we require weak localization  (WL) corrections (neglected
here) to be small, i.e.\  $ L < \xi$, where $\xi$ is the localization
length. In the presence  of time reversal (GOE) symmetry  $ \xi_o \sim
\ell \exp g_0 $ \cite{LernerImry}, and these two inequalities  are only
marginally compatible. On the other hand,  for a broken (e.g., due to an
applied magnetic field) time reversal (GUE) symmetry $\xi_u \sim \ell \exp
g_0^2 $, i.e.\ there exists a regime for which both Eq.\ (\ref{l1}) and
the requirement on the smallness of WL corrections are satisfied.

\begin{figure}\centering
\epsfxsize7cm\epsfbox{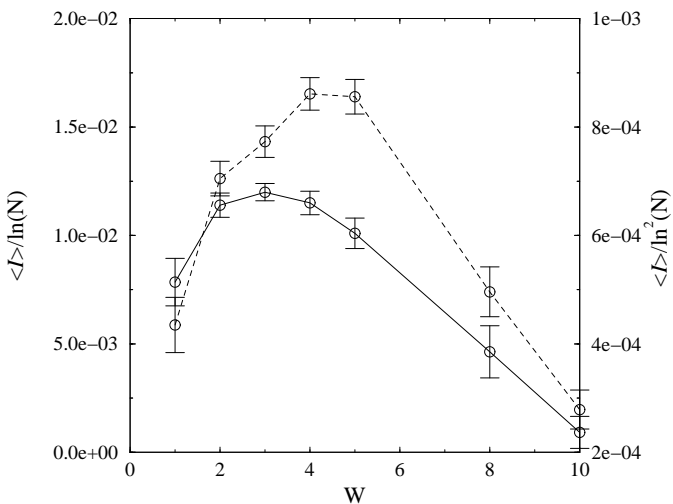}
\parbox{8.5cm}{\caption[]{\small
$\langle \langle {\cal I} \rangle_{disorder}/ \ln N  \rangle_{N}$
(solid line) and $\langle \langle {\cal I} \rangle_{disorder}/
\ln^2 N  \rangle_{N}$(dashed line)  vs. the disorder $W$ for an
$L=32$ system around quarter filling.} \label{fig.0}}\end{figure}

Arguably  the most intriguing  result concerning   $\left<{\cal I}\right>$
is the  dependence on disorder. It is clear from Eq.\ (\ref{i2}) that in
the weak disorder limit (and finite impurity range) the exponential
suppression  of the overlap is enhanced as $g$ decreases. On the other
hand one expects the AOC to vanish in the strongly localized limit. This
implies a {\it nonmonotonic} dependence on disorder. Such a behavior is,
indeed, found numerically (see Fig. \ref{fig.0}).

The analysis outlined above concerns with the {\it average} value of $\cal
I$. As we show below, the  distribution function   $\rho(\chi)$ is
remarkably broadly  shaped, so that the average alone does not provide us
with the full information on the behavior of $\chi$.

Our numerical study addressed a tight-binding Hamiltonian in the presence
of diagonal disorder: the site energies assumed a box distribution of
width $W$. The nearest-neighbor hopping, $t$, was chosen to be constant in
magnitude, but with a phase factor when a magnetic field is applied. The
underlying lattice was a two-dimensional square of area $L^2$ with
periodic boundary conditions. For each realization of onsite energies the
Hamiltonian is diagonalized exactly, both in the absence and in the
presence of an additional impurity, parametrized according to Eq.\
(\ref{imp}). The calculation is performed for up to $5\times10^4,
3.2\times10^4, 2\times10^4, 1.6\times10^4, 1.2\times10^4$ different
realizations for the sample sizes $L=16,20,24,28,32$, respectively, and
several values of disorder.

\begin{figure}\centering
\epsfxsize7cm\epsfbox{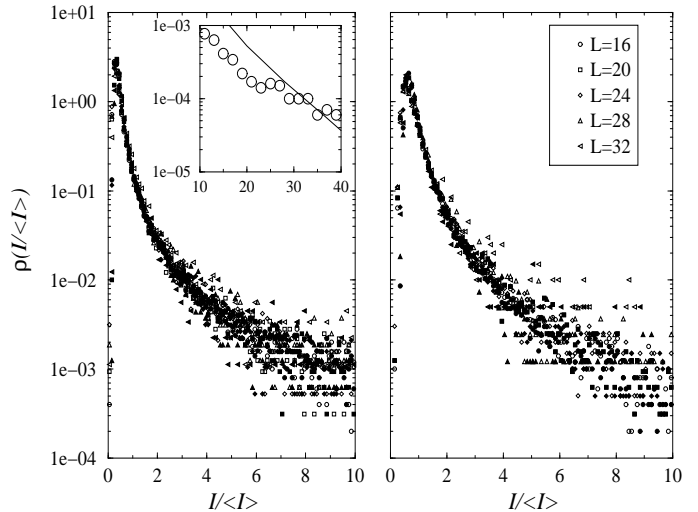}
\parbox{8.5cm}{\caption[]{\small
The distribution of ${\cal I}$ normalized by its average for different
system sizes $L$ at quarter filling $N=L^2/4$. Open symbols correspond to
$W=2t$ while filled ones to $W=3t$. Left panel: No magnetic field
(corresponding to GOE statistics). Right panel: Finite magnetic field (GUE
statistics). Note that the larger values of ${\cal I}/\langle{\cal
I}\rangle$ correspond to smaller values of $\chi$. Inset: The tail region
for the GOE statistics at quarter filling, $L=16$, $W=3t$. The line
corresponds to the two-level theory. } \label{fig.1}}\end{figure}

The broad  distribution of ${\cal I}$ (normalized by its average
value $\langle {\cal I}\rangle$) is presented in  Fig.\
\ref{fig.1} for different values of $L$ with filling $N/L^2=1/4$
and for $W=3t$ and $W=2t$. Here we consider an impurity with
$\lambda=1$ and $b=2$. The distribution we find is quite
remarkable. Clearly it is  even broader than
log-normal\cite{starkdifference}. The extended tail of the
distribution corresponds to small values of ${\chi}$ (large values
of ${\cal I}/\langle {\cal I} \rangle$) and gives the appearance
of power law behavior. The range of the tail shown in Fig.
\ref{fig.1} (before the statistical scatter of data points washes
it out) is at least $15$ standard deviations. Furthermore the
distribution seems to be governed by a single scaling parameter
${\cal I} /\langle {\cal I} \rangle$, and does not depend on
system size or on disorder. The distributions exhibit extended
tails and obey the single parameter scaling both for the GOE and
the GUE symmetries. Upon close inspection it appears that the
details of the tail differ between the GOE and the GUE scenarios,
the occurrence of atypical events been suppressed in the latter (a
"smaller tail"). This is accounted for below.

The clear deviation of the distribution's tail from a log-normal
shape suggests that large fluctuations in ${\cal I}$ are a
manifestation of few-level physics. This is  supported by a direct
diagrammatic estimation of fluctuations. The leading contributions
to $\left< {\cal I}^2\right>$ and to the higher cumulants of $\cal
I$ come from the same diagrams as those for the local density of
states (LdoS), as is clear from Eq.\ (\ref{I}). The LDoS
distribution by itself is log-normal \cite{IVL:88}; the additional
energy denominator in Eq.\ (\ref{I}) leads to  powers of (still)
logarithmic divergence in all orders of perturbation theory for
the cumulants of $\cal I$ higher than those for the LdoS
cumulants. This indicates that the distribution of $\cal I$ is
even wider than log-normal. Although it makes a direct
diagrammatic (or renormalization group) analysis of the
fluctuations of $\cal I$ hardly possible, this infrared divergence
indirectly indicates the importance of just a few energy levels
near the Fermi energy.

\begin{figure}\centering
\epsfxsize7cm\epsfbox{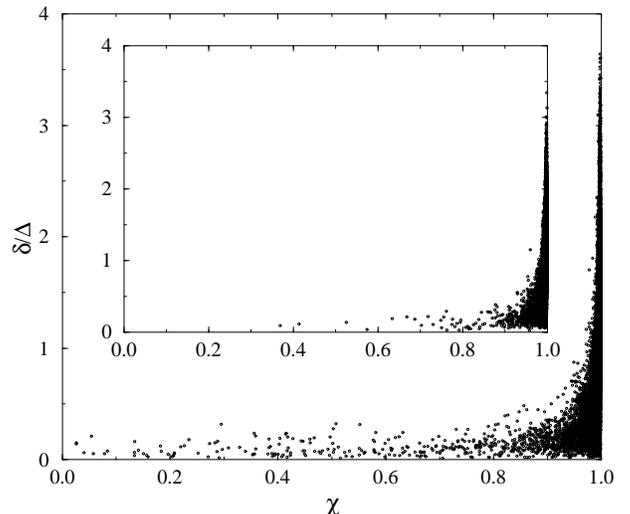}
\parbox{8.5cm}{\caption[]{\small
Scatter plot of $\chi$ vs. $\delta$ for $L=16$, $W=3t$ at quarter
filling with GOE statistics. Inset: GUE statistics.}
\label{fig.3}}\end{figure}

The final confirmation of the few level picture is provided by a
scatter plot of $\chi$ vs. the levels spacing $\delta$ between the
$N+1$st and $N$th level for the quarter filled case, Fig.
\ref{fig.3}. In these plots the differences (GOE vs. GUE) hinted
to in the distribution plots become much clearer. While there is a
finite (small) probability of finding small Anderson overlaps
($\chi < 1/2$) for the GOE case, there is almost no probability of
finding such small overlaps in the GUE case. At the same time
small values of $\chi$ are strongly correlated with small level
spacings, leading to suppression of the tail of $\rho$ in the GUE
case.

Since the level spacing at the Fermi energy is strongly correlated
with small values of the Anderson integral, it is reasonable to
try to develop an analytic description for the tail of the
distribution from the behavior of a two-level system representing
the last occupied and first unoccupied levels in the system
\cite{Vallejos}. Prior to the introduction of the impurity the
levels have the energy $-\delta/2$ and $\delta/2$ and wave
functions $\psi_N(\vec r)$ and $\psi_{N+1}(\vec r)$ respectively.
After the introduction of an impurity $\lambda \delta(\vec r=0)$,
the two level system is represented by the following Hamiltonian:
\begin{equation}
H=\left(\matrix{\delta/2 + \lambda |\psi_{N+1}(0)|^2 &
\lambda \psi_{N+1}^*(0)\psi_N(0) \cr
\lambda \psi_{N+1}(0)\psi_N^*(0) &
-\delta/2 + \lambda |\psi_N(0)|^2 \cr}
\right).
\label{2lev}
\end{equation}
One can bound $\chi$ from above
by $\chi'$, the contribution to the Anderson overlap due to the $N$-th
level (before and after the introduction of the impurity).
This is readily done by
diagonalizinig the above Hamiltonian:
\end{multicols}
                    \vspace*{-3.5ex} \hspace*{0.491\hsize}
            \begin{minipage}{0.48\hsize}$\,\!$\hrule
             \end{minipage}
\begin{eqnarray}
\label{overlap} \chi'=\frac12 + \frac12\,\frac{\delta  +
\lambda(|\psi_{N+1}(0)|^2-|\psi_N(0)|^2) } {\sqrt{(\delta/2 +
\lambda(|\psi_{N+1}(0)|^2-|\psi_N(0)|^2)/2)^2 + \lambda^2
|\psi_{N+1}(0)|^2 |\psi_N(0)|^2}}\,.
\end{eqnarray}
\vspace*{-3.5ex}
            \begin{minipage}{0.48\hsize}$\,\!$\hrule
             \end{minipage}\vspace*{1.5ex}
                  \begin{multicols}{2}
We further assume that the distribution of the level spacing
$\delta$ is given by
$$P(\delta/\Delta)=\begin{cases}{(\pi \delta/2
\Delta)\exp[-(\pi/4)(\delta/\Delta)^2],&  GOE  \cr (32/\pi^2
(\delta/\Delta)^2)\exp[-(4/\pi)(\delta/\Delta)^2],&
 GUE}
\end{cases}$$
 and the wave function distributions are $P(\Re{\rm e}\,
\psi(0))=P({\Im{\rm m}}\, \psi(0))=(\widetilde N/2 \pi)^{1/2}
\exp(-\widetilde N \Re{\rm e}\, \psi(0)^2/2)$ for the GUE case,
while in the GOE case there are only real components of the
wavefunction ($\widetilde N\sim (p_F L)^2$). The distribution of
$\chi'$ is given then by
\begin{eqnarray}
\label{distri} \rho (\chi) = \int_0^{\infty} \!\dd \delta\
P\!\left(\frac\delta\Delta\right) \int  \text{D}\psi_{N+1}\text{D}
\psi_N\, \delta(\chi'-\chi),
\end{eqnarray}
where $\chi'$, which depends on all the integration variables, is
given by Eq.\ (\ref{overlap}), and $\text{D} \psi_{N(N+1)}$
denotes an integration over the real and imaginary parts of
$\psi_{N(N+1)}(0)$. The integration is performed numerically using
a Monte Carlo method \cite{tl}, and the results for the tail
region are shown in the inset of Fig. \ref{fig.1}. As can be seen,
deep in the tail region $\rho (\chi)$ fits the distribution
reasonably well.

To summarize, we have found three main peculiarities in the
physics of AOC in  diffusive  2d systems. First,  the average
value of the overlap between the original and the modified ground
states decreases with $N$ faster than any power law. Second, it
shows a non-monotonic dependence on disorder. Finally, the
logarithm of the overlap shows remarkably wide fluctuations with
disorder, with long tails which are mainly due to the
contributions from only a few levels near the Fermi energy.

We acknowledge discussions with  C. Lewenkopf and R.O. Vallejos.
This work was supported in part by the U.S.-Israel BSF, the GIF
foundation, the ISF-Centers of Excellence Program, the Minerva
foundation, and by the Leverhulme trust. Y.G., R.B. and I.V.L.
acknowledge the kind  hospitality extended to them at Princeton
University and NEC.

%\vskip -1truecm

\vspace*{-3cm}

\end{multicols}


\begin{thebibliography} {50}
\bibitem {corrections}
There are corrections to the results (\ref{i2}) and (\ref{i1}) involving
terms $\sim \ln k_F b$ and $ \sim \ln g_0$ small compared to $\ln N$.
\bibitem{footnote_e} Note that since
$\langle G^{R/A} \rangle$ is essentially energy independent we omit the
dependence on $\varepsilon,\varepsilon'$. More precisely, the $k_F$
factors in Eq.\ (\ref{gg}) should be replaced by $k(\varepsilon)$ or
$k(\varepsilon')$ in the energy integrals. The $J_0^2$ or the $\sin^2$
terms are then modified into oscillatory terms which effectively restrict
the energy integrals to $|\varepsilon - \varepsilon'| \le
\varepsilon_F/k_F b$. We end up with the $\ln N$ factors being replaced by
$\ln (N/k_F b)$ \cite{corrections}.
\bibitem {long-ranged} Note that the long-range contribution
arises from an energy shell $\sim D/b^2 \le \hbar/\tau$ around the Fermi
energy in contrast to the original AOC or the short range term, which are
the entire Fermi sea contributions.
\bibitem{starkdifference} By contrast the distribution of
$1/{\cal I}$ , for which the tail probes values of $\chi$ close to one,
is clearly log-normal and depends on the disorder.
\bibitem {tl} One can simplify Eq.\ (\ref{distri}) to the level of a
double integral which can then be computed numerically. A direct numerical
evaluation of the integral Eq.\ (\ref{distri}) is faster.
\end{thebibliography}
\end{document}